\documentclass[twocolumn]{aastex63}

\newcommand{\msunyr}{\ensuremath{\mathit{M}_{\odot}{\rm yr}^{-1}}}   
\newcommand{\msun}{\ensuremath{\mathit{M}_{\odot}}}   
\newcommand{\lsun}{\ensuremath{\mathit{L}_{\odot}}}                  
\newcommand{\rsun}{\ensuremath{\mathit{R}_{\odot}}}                  

\newcommand{\lstar}{\ensuremath{\mathit{L}_{\star}}}                 
\newcommand{\mdot}{\ensuremath{\dot{M}}}                             
\newcommand{\teff}{\ensuremath{\mathit{T}_{\rm eff}}}                
\newcommand{\vinf}{\ensuremath{\upsilon_{\infty}}}                          
\newcommand{\K}{\ensuremath{\mathrm{K}}}                 

\usepackage{amsmath}	        
\newcommand{\mcocore}{\ensuremath{M_{\operatorname{CO-core}}}}  
\newcommand{\fms}{\ensuremath{f_\mathrm{MS}}}                             
\newcommand{\mdotlbv}{\ensuremath{\dot{M}_{\rm LBV}}}                             

\definecolor{orange}{rgb}{0.726, 0.015, 0.015}
\newcommand\E[1]{{\color{orange} \bf #1}} 

\submitjournal{ApJ}

\shorttitle{BH mass regulated by LBVs}
\shortauthors{Groh et al.}

\begin{document}

\title{Massive black holes regulated by luminous blue variable mass loss and magnetic fields} 

\correspondingauthor{Jose H. Groh}
\email{jose.groh@tcd.ie}

\author{Jose H. Groh}
\affiliation{Trinity College Dublin, The University of Dublin, Dublin, Ireland }

\author{Eoin J. Farrell}
\affiliation{Trinity College Dublin, The University of Dublin, Dublin, Ireland }

\author{Georges Meynet}
\affiliation{Department of Astronomy, University of Geneva, Maillettes 51, 1290, Versoix, Switzerland}

\author{Nathan Smith}
\affiliation{Steward Observatory, University of Arizona, 933 N. Cherry Avenue, Tucson, AZ 85721, USA}

\author{Laura Murphy}
\affiliation{Trinity College Dublin, The University of Dublin, Dublin, Ireland }

\author{Andrew P. Allan}
\affiliation{Trinity College Dublin, The University of Dublin, Dublin, Ireland }

\author{Cyril Georgy}
\affiliation{Department of Astronomy, University of Geneva, Maillettes 51, 1290, Versoix, Switzerland}

\author{Sylvia Ekstr\"om}
\affiliation{Department of Astronomy, University of Geneva, Maillettes 51, 1290, Versoix, Switzerland}

\begin{abstract}
We investigate the effects of mass loss during the main-sequence (MS) and post-MS phases of massive star evolution on black hole (BH) birth masses. We compute solar metallicity Geneva stellar evolution models of an 85 \msun\ star with mass-loss rate (\mdot) prescriptions for MS and post-MS phases and analyze under which conditions such models could lead to very massive BHs. Based on the observational constraints for \mdot\ of luminous stars, we discuss two possible scenarios that could produce massive BHs at high metallicity. First,  if a massive BH progenitor evolves from the observed population of massive MS stars known as WNh stars, we show that its average post-MS mass-loss rate has to be less than $1\,\times10^{-5}\,\msunyr$. However, this is lower than the typical observed mass-loss rates of luminous blue variables (LBV). Second, a massive BH progenitor could evolve from a yet undetected population of $80-85$ \msun\ stars with strong surface magnetic fields, which could quench mass loss during the evolution. In this case, the average mass-loss rate during the post-MS LBV phase has to be less than $5\,\times10^{-5}\,\msunyr$ to produce 70 \msun\ BHs. We suggest that LBVs that explode as SNe have large envelopes and small cores that could be prone to explosion, possibly evolving from binary interaction (either mergers or mass gainers that do not fully mix). Conversely, LBVs that directly collapse to BHs could have evolve from massive single stars or binary-star mergers that fully mix, possessing large cores that would favor BH formation. 
\end{abstract}


\section{Introduction} \label{intro}

Massive stars have a key impact throughout the history of the Universe, being the main contributors to the emission of ionizing photons, energy and production of some chemical elements.  The majority of massive stars leave a neutron star (NS) or black hole (BH) as compact remnant \citep[e.g.][]{maeder_araa00,langer12}, which can be observed with electromagnetic radiation \citep[e.g.][]{casares14} and gravitational waves \citep[e.g.][]{abbott16}. The most massive stars may be affected by the pair-creation instability, which can produce either pulsations and strong mass loss that still leaves a BH remnant, or a total disruption of the star in a pair-instability supernova with no remnant left \citep[e. g.][]{woosley17}. The mass of the Carbon-Oxygen core (\mcocore) at the end of the evolution is thought to be one of the key parameters that set the final fate of a massive star \citep[e.g.][]{heger00,heger03,woosley17}.

Current observational evidence of BH detections suggest a maximum BH mass of around 30 \msun\ at solar metallicity \citep[e.g.][]{abbott16, zampieri09, belc10, spera15, belc16a}. Stellar evolution models of single stars also predict a maximum black hole mass of $\sim 20 -30~\msun$ at solar metallicity \citep[e.g.][]{groh19} when usual assumptions are made regarding mass loss, rotation, and convective core properties.

Detecting massive BHs help to shed light on the processes that operate during stellar evolution, and several ongoing spectroscopic surveys have the potential to increase the sample of detected galactic BHs. Recently, a very massive BH has been proposed to orbit a B-type star in the outskirts of the Milky Way \citep{liu19}. These authors favor that the LB-1 system consists of a BH with a mass of  $68^{+13}_{-11}$ \msun\ and a B3V star with mass of $8^{+1.2}_{-0.9}$~\msun. They suggest an orbital period of $P_\mathrm{orb}= 78.9 \pm 0.3$~d. Following the initial discovery of the binary system, several studies have put into question the existence of such a massive black hole in LB-1 \citep{abdulmasih20,elbadri20,eldridge20,irrgang20,simondiaz20}, and the debate is still ongoing \citep{liu20}. 

To explain a potentially massive BH at solar metallicity, \citet{belczynski19} proposed a reduction of stellar wind mass-loss rates by a factor of 3 -- 5 throughout the entire evolution. These authors can explain the formation of a 70~\msun\ BH as evolving from a non-rotating star of initial mass 85 \msun with such reduced mass-loss rates if the progenitor directly collapses to a BH without losing significant mass in the process. In this case, the progenitor would have $\mcocore=27.6~\msun$ and narrowly avoid pulsational pair-instability, which is thought to occur for CO cores in the range $\mcocore=28 - 54~\msun$ \citep{woosley17}. However, the Roche-lobe radius of the BH is around 200 \rsun and this star would expand to a radius of $\sim 350~\rsun$, larger than the binary orbit, during the evolution. \citet{belczynski19} also discuss the effects of rotation on the progenitor core properties.

Here we investigate the available observational data for luminous massive stars in the upper part of the Hertzsprung-Russell (HR) diagram that could be the progenitors of massive BHs. We then compute numerical stellar evolution models with the Geneva code to study the mass budget of progenitors of massive BHs. Our results indicate that the uncertain mass loss by luminous blue variable stars (LBVs) has a key impact in the black hole birth mass function.

\section{\label{obs}How much can mass loss be reduced based on observations?}
To produce a $\sim70~\msun$ BH like the one that was originally proposed to exist in LB-1, its progenitor would have evolved from a star that had an initial mass of around 85 \msun, or acquired a similar mass at some point during its evolution by interacting with a companion \citep{belczynski19}. Stellar evolution models indicate that stars more massive than 85~\msun\ have luminosities $ \log \lstar/\lsun \gtrsim 6.1$ \citep{ekstrom12}. Therefore, in this section, we focus on observational constraints for mass loss in these very luminous stars. We refer the reader to \citealt{smith14araa} for a broader review of mass loss in massive stars and \citet{vink15} for a review on the impact of very massive stars in the local Universe.

\begin{figure}[ht!]
\center
\resizebox{0.475\textwidth}{!}{\includegraphics{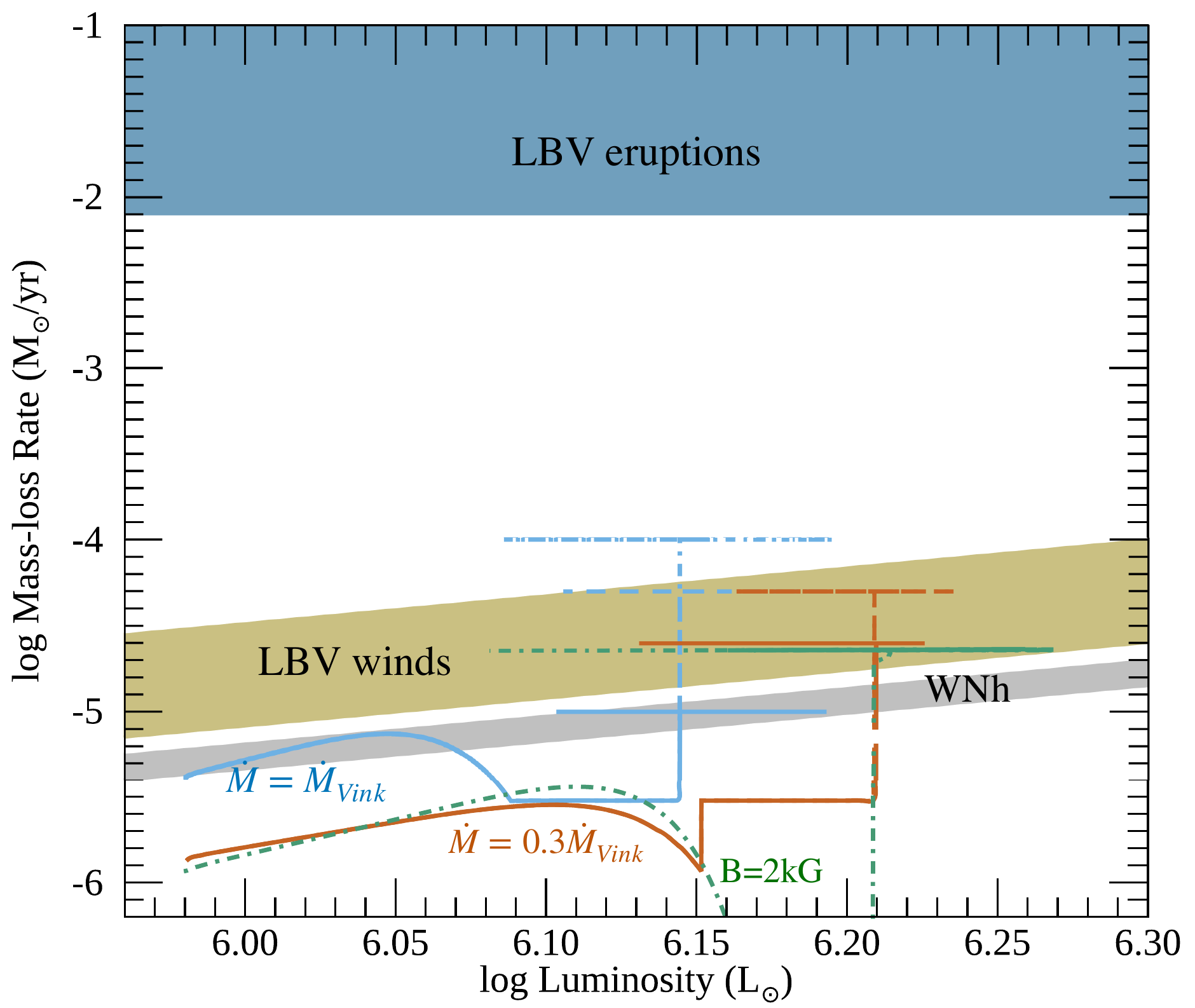}}
\caption{Mass-loss rates of luminous stars derived from observations of WNh, and LBV stars. The data were compiled from \citet{martins07,martins08}, \citet{clark12b}, \citet{smith14araa} and references therein. We also overplot our Geneva stellar evolution models for different MS (indicated by the labels) and post-MS mass-loss rate prescriptions, in addition to a Geneva model including a surface dipolar magnetic field of 2 kG (dot-dashed green line).  \label{fig1}}
\end{figure}

MS stars at these luminosities are predominantly observed with emission-line spectra of the WNh sub-type, typically with $\mdot \gtrsim 5\times 10^{-6} \msunyr$ derived using detailed spectroscopic modelling \citep{martins08,martins13,crowther10,bestenlehner14,smith14araa}. These observed mass-loss rates are broadly in agreement the theoretical prescription from \citet{vink01}. An example is NGC3603-A1b, for which the derived \mdot\ from CMFGEN spectroscopic modelling is actually slightly higher than the theoretical \citet{vink01} prescription \citep{crowther10}. Further support for the relatively high mass-loss rates of MS stars is provided by theoretical work on their spectral morphologies. Lower mass-loss rates would lead to O-type absorption line spectra, which is inconsistent with the morphology of luminous WNh stars \citep{crowther10,martins17}. In addition, \citet{vink12} derived a theoretical framework for a mass-loss rate that characterizes the transition between optically-thin and optically-thick winds. Such a transition is around $\mdot \sim 10^{-5} \msunyr$ at solar metallicity. Lastly, there is no evidence that WNh stars have strong magnetic fields to confine the wind and reduce the effective mass-loss rate, as has been proposed for lower-mass and less luminous O-type magnetic stars \citep{petit17,zsolt19}. In fact, because WNh are characterized by strong mass loss ($\mdot \gtrsim 10^{-5} \msunyr$; \citealt{martins08}), they seem to be incompatible with strong magnetic quenching of mass loss. Based on this observational evidence, we do not see room for an outright reduction of mass-loss rates by factors of 3-5 during Hydrogen burning for 85~\msun~stars if they are to produce very massive BHs, unless the progenitor evolved from a strongly magnetized luminous star that is unlike the observed WNh population. Because of their lower \mdot, these stars should be spectroscopically similar to O or Of stars.

At lower temperatures, the upper HR diagram for stars with $ \log \lstar/\lsun \gtrsim 6.0$ is dominated by LBVs and blue hypergiants. These stars have spectra characterized by P-Cyg profiles and emission lines, which indicate high \mdot. Detailed spectroscopic analyses using the radiative transfer code \textsc{CMFGEN} have shown that luminous LBVs do indeed have typical mass-loss rates of $\mdot  > 10^{-5}~\msun/yr$ \citep{ghd09,ghd11,clark12b}. For example, the prototypical LBV AG Carinae, which is in the luminosity range through which 85~\msun\ stars evolve, has a current mass around 60-70~\msun\ \citep{ghd11} and a quiescent stellar-wind mass loss that varies between $\mdot  \simeq 1.5  \times 10^{-5}$ and  $\sim 1.0  \times 10^{-4}~\msunyr$ \citep{ghd09,stahl01}. Thus, using average post-MS mass losses $\mdot  < 10^{-5} \msunyr$ would be inconsistent with the LBV observations. On top of quiescent stellar winds, LBVs also show eruptive mass loss episodes \citep{so06}, which would only add to the total amount of mass lost by an LBV. Because the frequency of LBV eruptions and the mass lost per eruption are unknown, stellar evolution models have to rely on average mass-loss rates of LBVs, which is what we include in our models below.

\begin{figure}[ht!]
\center
\resizebox{0.475\textwidth}{!}{\includegraphics{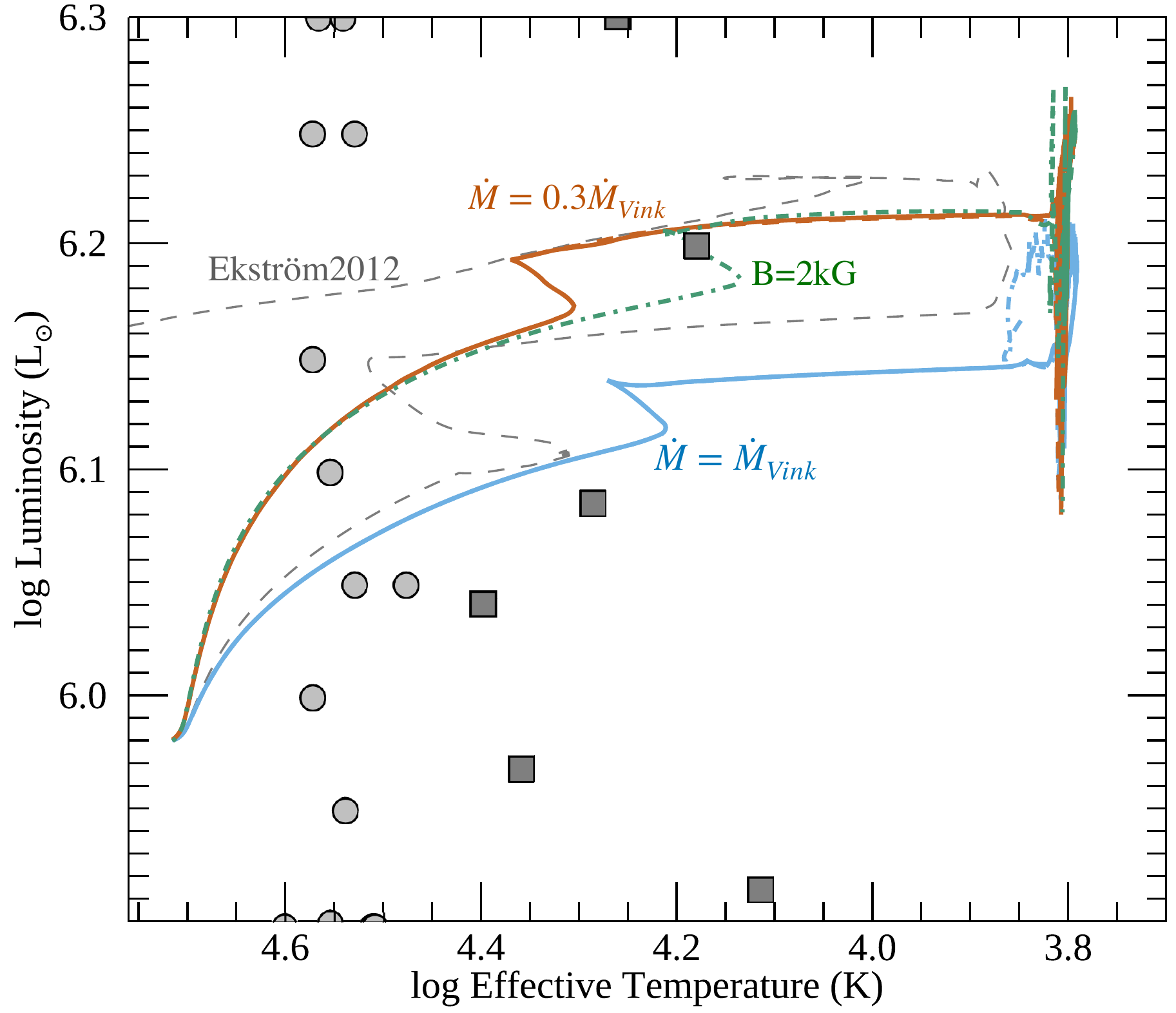}}
\caption{HR diagram of our Geneva stellar evolution models, showing the set of models with $\fms=0.3$ (orange) and $\fms=1.0$ (blue).  A model including mass-loss quenching by a surface dipolar magnetic field of $B=2$~kG is shown in green. All models spend their post-MS phase as LBVs at $\log \teff \simeq 3.8$, and the different colors at that point indicate the different values of \mdotlbv\ adopted by our models. We also plot the luminosity and effective temperature derived from observations of WNh (light gray circles), and LBV (candidate) stars (dark grey squares). The observational data was compiled from \citet{martins07,martins08}, \citet{clark12a}, \citet{smith19} and references therein. \label{fig2}}
\end{figure}\

\section{\label{models}Geneva stellar evolution models for different LBV mass-loss rates}

\begin{table}
\centering
\caption{Summary of stellar evolution models with an initial mass of $85 \msun$ and different mass loss rates during core H burning and the LBV phase. We list the mass loss scaling factor applied during core H burning (\fms), the constant mass loss rate applied during the post-MS evolution (\mdotlbv), the mass lost (in \msun) during the MS evolution, the mass lost during the post-MS evolution and the total stellar mass at the end of the evolution (M$_{\rm final}$)}
\label{table1}
\begin{tabular}{rrrrr}
\hline
\fms & \mdotlbv / 10$^{-5}$ & \msun\ lost   & \msun\ lost & M$_{\rm final}$\\
& \msunyr\ &  (MS) & (post-MS)  & \msun\ \\
\hline
1.0 & 1.0 & 16.4 & 3.4 & 65.2~\msun\ \\
1.0 & 5.0 & 16.4 & 16.4 & 52.2~\msun\ \\
1.0 & 10.0 & 16.4 & 33.9 & 34.7~\msun\ \\
\hline
0.3 & 2.5 & 6.4 & 7.9 & 70.7~\msun\ \\
0.3 & 5.0 & 6.4 & 15.9 & 62.7~\msun\ \\
0.3 & 10.0 & 6.4 & 31.7 & 46.9~\msun\ \\
\hline

\end{tabular}

\end{table}

Based on these observational constraints for minimum mass-loss rates for Hydrogen-burning stars and LBVs, we compute numerical stellar evolution models with an initial mass of 85 \msun\ using the Geneva code. A star with an initial mass of 85 \msun\ could lead to the most massive BHs that could be formed at solar metallicity, producing final CO core masses just below the pulsational pair-instability regime \citep{belczynski19}.

We summarize our main assumptions below and refer the reader to \citet{ekstrom12}, \citet{georgy13} and \citet{georgy17} for further details on the Geneva code. Our models have solar metallicity ($Z=0.014$), with solar abundances values from \citet{asplund09}. To isolate the effect of mass loss on the evolution final masses of massive stars, we focus on non-rotating models only. As a consequence, we avoid the two main feedback effects of rotation on mass loss. Firstly, rotating models are more luminous than a single star model of same initial mass, which in turn leads to higher values of \mdot. Secondly, high rotation may produce rotationally-induced mass loss, although the magnitude of such effect is under debate \citep{mm_omega00,muller14}. We terminate the stellar evolution models at the end of core He or Carbon burning, depending on numerical convergence issues.
 
Our models use the \citet{vink01} prescription for \mdot\ for the core Hydrogen burning phase, which we define here as when the central H abundance is greater than $10^{-4}$. We compute several models with the \citet{vink01} prescription scaled by factors of $\fms = \mdot/\mdot_\mathrm{Vink} = 1.0 $, 0.3, and 0.0. Models with $\fms=1.0$ would roughly match the mass-loss rates in the observed WNh population, while $\fms= 0.3$ and 0.0 mimic the quenching of mass-loss by magnetic fields.  For comparison, we also compute a model with a surface dipolar magnetic field of 2 kG, following the implementation of mass-loss quenching by magnetic fields from \citet{georgy17}, which is based on the equations from \citet{petit17}. We suppress the 10x increase in \mdot\ that is included in the \citet{vink01} prescription when the star crosses the bi-stability jump. Models including the bi-stability jump would produce lower BH masses. Instead, we reduce the jump in \mdot\ to prevent strong mass loss at the end of the MS that would dominate the mass budget \citep{gme14}, as included in e.g. \citet{ekstrom12}. We also refer the reader to \citealt{zsolt17} for further discussions on the impact of mass loss and the bi-stability jump on massive star evolution.

For the post-MS phase, we switch to an average LBV mass loss (\mdotlbv) when $\teff\lesssim15000~\K$. The exact value of \teff\ at which we switch prescriptions has little impact on the total mass lost. We compute models for $\mdotlbv = 1.0, \,2.5, \,5.0$, and $10.0  \times 10^{-5}~\msunyr$. Values of \mdotlbv\ much greater than $10^{-4}~\msunyr$ would remove the entire H envelope and produce a Wolf-Rayet star with a final mass of around 20~\msun\ \citep{gme13}.
 
Figure \ref{fig2} shows our evolutionary tracks in the HR diagram. The model with $\fms = 0.3$ (blue) has a higher luminosity than the model with $\fms = 1.0$ (orange) at all times due to its higher mass. Both models evolve through the region of the HR diagram where WNh stars are observed. During the post-MS, the models are spectroscopically similar to LBVs (\citealt{gme14}) due to the high value of \mdot. This is expected to occur as the star evolves and becomes closer to the Eddington limit. If the mass-loss rates are low enough, the star retains its H envelope and remains as an LBV until the end of its evolution.

\section{\label{massbudget} The final mass of massive stars} 

\begin{figure}[ht!]
\center
\resizebox{0.475\textwidth}{!}{\includegraphics{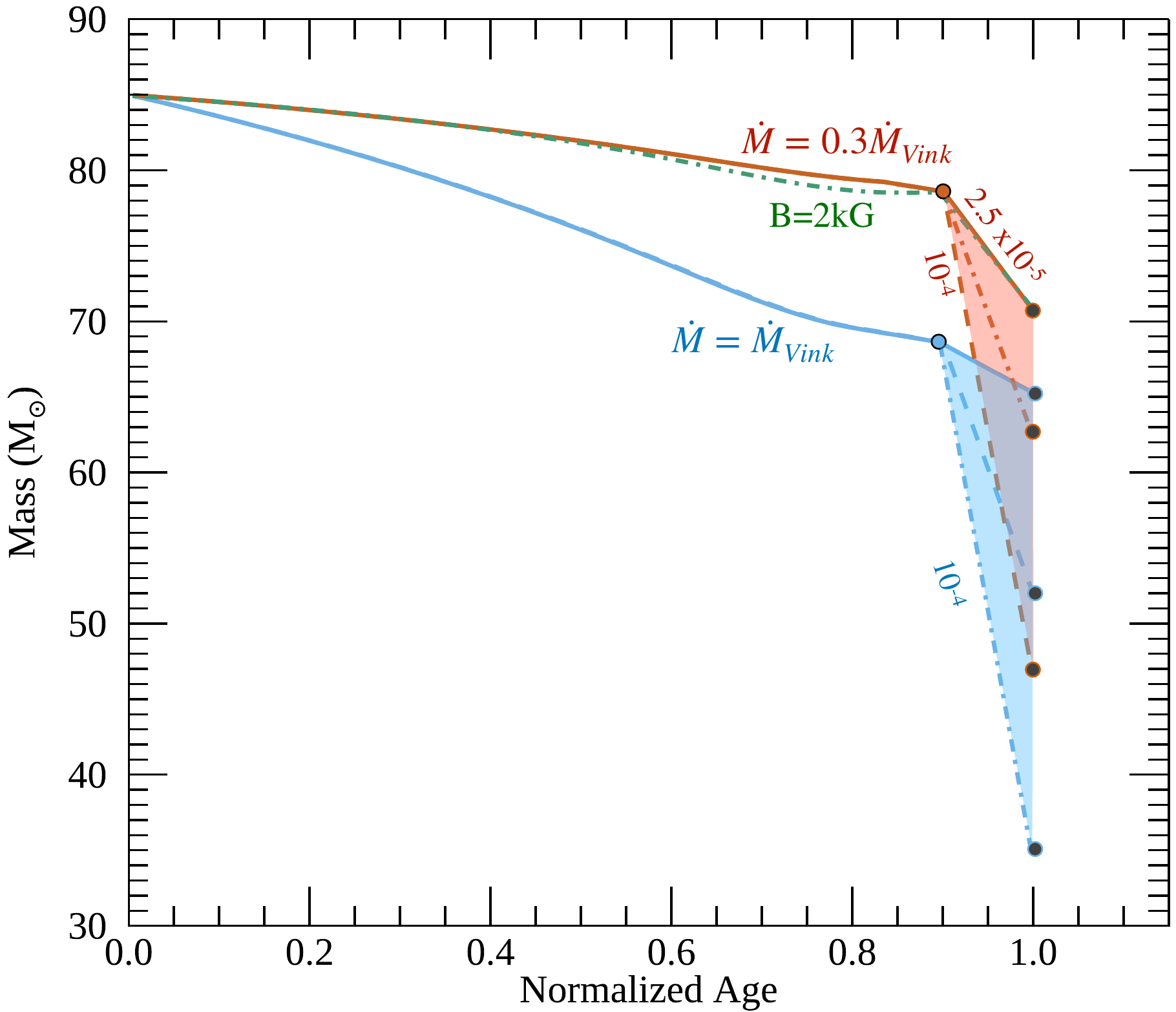}}
\caption{Evolution of the stellar mass as a function of the normalized age for our Geneva stellar evolution models with different mass-loss rate prescriptions. The labels indicate the values of the MS mass-loss rate (blue and orange), the surface dipolar magnetic field strength when applicable (green) and \mdotlbv\ of our models. \label{fig3}}
\end{figure}

Figure \ref{fig3} presents the evolution of the total stellar mass a function of normalized age for our models with different MS and post-MS mass loss prescriptions. For now we focus on the 85 \msun\ models with no convective core overshooting,  but similar conclusions are reached for stars in the range 40--100~\msun\ since they also go through an LBV phase \citep{ekstrom12, gme14}. Table~\ref{table1} summarizes the mass lost by our models in different evolutionary phases.

Models with $\fms=1.0$ correspond to the evolution predicted based on the mass-loss rates of the observed population of WNh stars in the Milky Way. These models reach the end of the MS with a mass of 68.6~\msun, losing 16.4~\msun. The star would need to lose mass during the post-MS LBV phase at an average rate of $\mdotlbv=1.0 \times 10^{-5}~\msunyr$ to finish its evolution as a 65~\msun\ BH, assuming no mass loss during fallback. This time-averaged LBV mass-loss rate is much lower than the observational constraints we discussed earlier (Fig. 1). For more realistic values of $\mdotlbv\ = 5.0  \times 10^{-5}~\msunyr$ and  $1.0  \times 10^{-4}~\msunyr$, the star would produce  52~\msun\ and 35~\msun~BHs, respectively. These are still relatively massive BHs compared to those detected by LIGO gravitational wave observations \citep{abbott19}. Our models indicate that the uncertain LBV mass loss leads to a wide range of final BH masses. For these reasons, obtaining a large observational sample BH masses from electromagnetic and gravitational waves could provided important constrains on the mass-loss history of massive stars. We conclude that to form very massive BHs  that evolve from the observed population of MS WNh stars, these objects have to lose mass at a much lower rate than those of observed luminous post-MS stars such as LBVs. However, if stars can avoid losing a large amount of mass during pulsations due to pair instability \citep{belczynski19}, it is possible for a $\sim 100 \msun$ star to have mass loss rates more similar to the observed luminous post-MS and still produce a $\sim 70 \msun$ BH.

The model with $\fms=0.3$ finishes the MS with a mass of 78.6 \msun, losing 6.4~\msun. Not surprisingly, this is similar to the MS mass loss in the model presented by \citet{belczynski19}, as they scale the mass-loss rates by a factor of $\fms=0.3$. Based on the available \mdot\ constraints of massive MS stars in this mass range, this model would correspond to the evolution of an unseen population of massive stars with relatively weak winds at high luminosity. They could correspond to strongly-magnetized $\sim 85~\msun$ stars, perhaps formed via mergers in a similar fashion as for $\sim15-20~\msun$ magnetic stars such as tau Sco \citep{schneider19,schneider20}.  Because of the strong surface magnetic fields, these stars would have their mass-loss rates quenched similarly to magnetic late-type O stars \citep{uddoula02,uddoula09,petit17,zsolt19}. Using the scaling for the mass-loss quenching (e.g. Eq. 3 from \citealt{petit17}), we find that a dipolar magnetic field strength of roughly around 2 kG would be required to quench the mass-loss rate to $\sim 30\%$ of what would be expected in the absence of surface magnetic fields. This assumes a stellar radius of 25~\rsun, $\mdotlbv\ = 1.1  \times 10^{-5}~\msunyr$ (before quenching), $\vinf=2200 km/s$  (before quenching), which are roughly the parameters expected for $85~\msun$ stars at the middle of the MS using stellar evolution models and the Vink \mdot\ prescription. Indeed, our detailed numerical model that includes a dipolar surface magnetic field of 2kG loses a similar amount of mass during the MS as the model with $\fms=0.333$.

Regardless of the MS evolution, mass loss during the post-MS evolution, possibly as an LBV, has strong impact on the BH birth mass function. Depending on \mdotlbv, these models with $\fms=0.3$ finish their evolution with a final mass of 75.3, 70.7, 64.4 and 52.0 \msun\, for mass loss rates of $\mdotlbv = 1.0, \,2.5, \,5.0$, and $10.0  \times 10^{-5}~\msunyr$ respectively (see Table \ref{table1}).

Under our assumption for the structure of the radiative envelope, the envelope is extended, as in to other numerical stellar evolution models \citep[e.g.]{choi16}. This means that our models do not solve the issue raised by \citet{liu19} and \citet{belczynski19} that the stellar radius of the progenitor of the putative LB-1 BH becomes larger than the size of the orbit. This can potentially be solved by accounting for a porous, structured radiative envelope which could potentially have a lower effective opacity \citep{owocki04}. Three-dimensional radiation hydrodynamic models would be needed to address this issue \citep{jiang18}.

\section{Implications for black hole birth masses} \label{imp}
 Based on the available observational constraints for \mdot\ of massive, luminous stars, we discuss two possible scenarios for the progenitor of 40--70\msun\ BHs at high metallicities. First, if the progenitor of these BHs volved from the observed population of WNh stars, it managed to lose mass with a maximum average rate of $1.0  \times 10^{-5}~\msunyr$ during its post MS phase. However, this value is lower than the typically observed mass-loss rates of LBVs. It is possible that some single stars might not go through the LBV phase as all and instead evolve as blue hyper-giants with lower mass-loss rates. 
 A second possibility is that the progenitor evolved from a yet undetected population of $\sim85~\msun$ stars with strong magnetic fields, perhaps originated in a merger event as proposed for O-type stars \citep{schneider19}. Surface magnetic fields could quench mass loss during the MS phase, as has been proposed for lower-mass O-type stars \citep{petit17,zsolt19} and very massive stars that produce pair-instability SN \citep{georgy17}. The strongly magnetized massive stars scenario could explain the recent claim from \citet{belczynski19} that reduced MS \mdot\ are needed for explaining the mass of the LB-1 BH that was originally proposed by \citet{liu19}. In this case, the average mass-loss rate during the post-MS LBV phase must be less than $5.0 \times 10^{-5}~\msunyr$. This is still a tight mass budget if we consider that observed luminous LBVs such as AG Carinae have average mass-loss rates from quiescent stellar winds close to or above this level. Eruptive mass loss that characterizes LBVs would add to the total mass lost.
 
In this paper we show that LBVs could be direct progenitors of massive black holes, which require a direct collapse and no SN explosion. This is apparently at odds with the evidence that LBVs are the direct progenitors of some SN events \citep{smith14araa}. To reconcile these two outcomes, we suggest that massive single stars (or mergers that fully mixed) evolve to have large cores, produce BH progenitors and perhaps avoid the LBV phase, while massive stars which form due to binary interaction (either mergers or mass gainers that do not fully mix; \citealt{smith15a}) may have smaller cores, larger envelopes, evolve through an LBV phase and could produce a SN event \citep{justham14}.

We encourage further searches for massive compact objects around galactic stars. Our main conclusion is that the formation of massive black holes at solar metallicity is possible depending on surface magnetic fields and post-MS mass loss. We find that the most massive black holes at solar metallicity could form from fallback in stars that retain a large H envelope, avoiding the WR phase. Before collapsing to BHs, these stars should resemble LBVs, i.e. unstable massive stars close to the Eddington limit. Some could show the characteristic S-Doradus type variability of LBVs. Improved observational constraints on the mass loss during the LBV stage is crucial to have a comprehensive view of black hole birth masses.

\bibliography{refs}{}
\bibliographystyle{aasjournal}

\end{document}